\documentclass[aps,prb,twocolumn,groupedaddress,showpacs,floatfix]{revtex4}

\usepackage{graphicx}
\usepackage{epsfig}
\usepackage{dcolumn}
\usepackage{bm}
\usepackage{amsmath}
\usepackage{verbatim}
\usepackage[usenames]{color}

\newcommand{\figa} {
\begin{figure}
\centering
\includegraphics[width=0.4\textwidth]{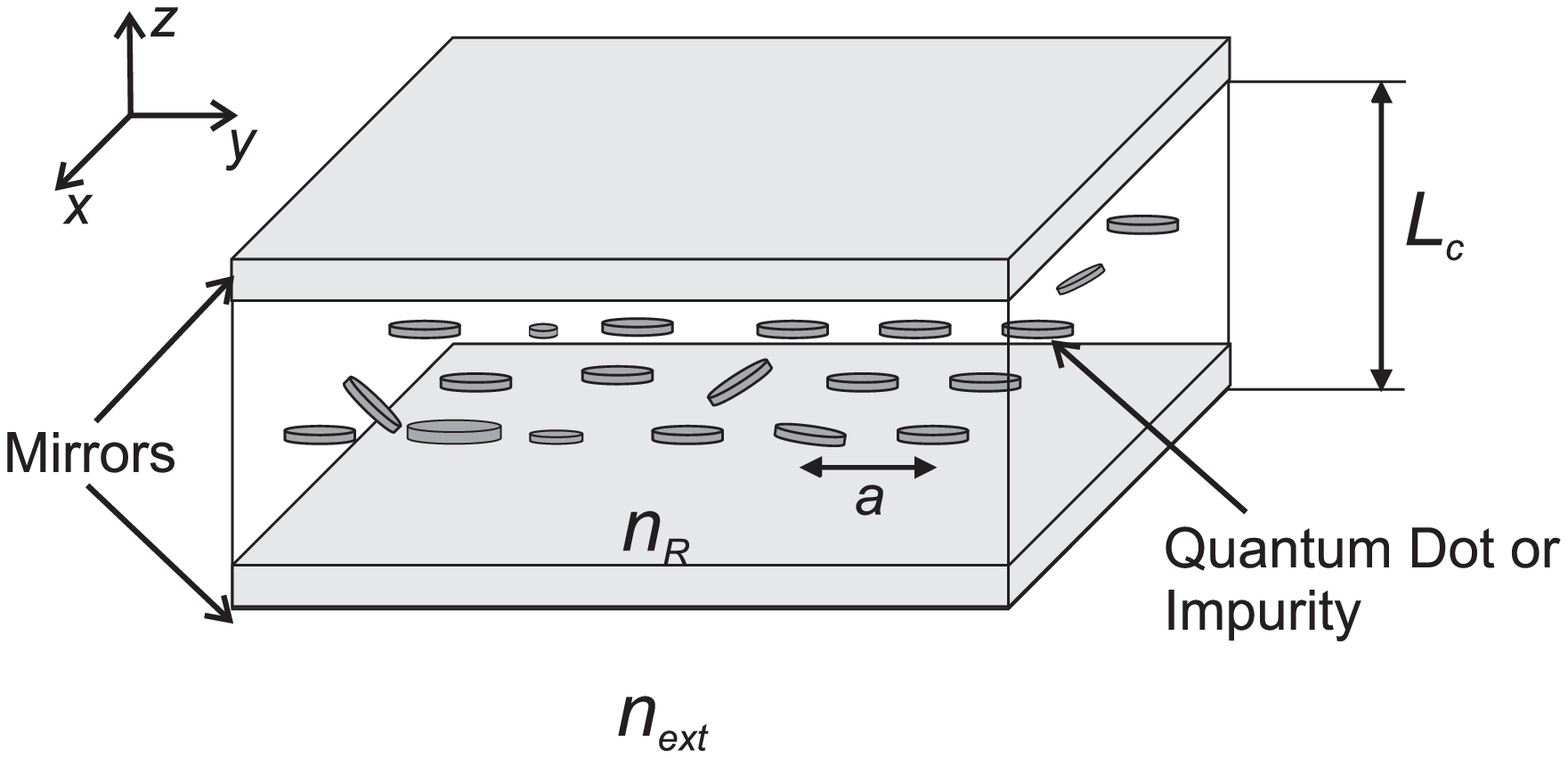} 
\caption{Scheme of the system studied. A lattice of quantum dots
or impurities is embedded in a planar microcavity. The energy,
position, and oscillator strength of the localized excitons fluctuates
from site to site.}
\label{figa}
\end{figure}
}

\newcommand{\figca} {
\begin{figure}
\centering 
\includegraphics[width=0.3\textwidth]{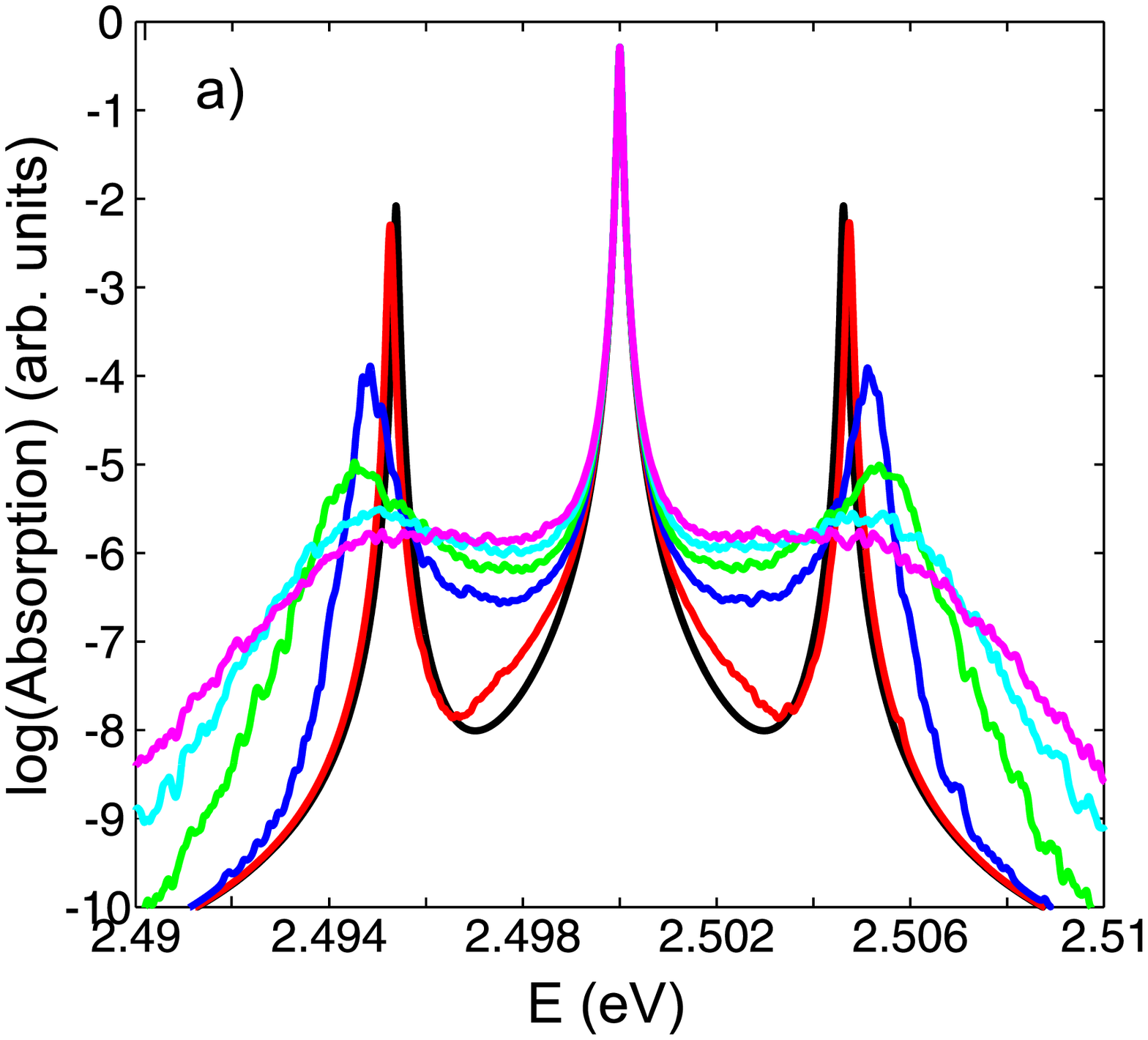}
\includegraphics[width=0.3\textwidth]{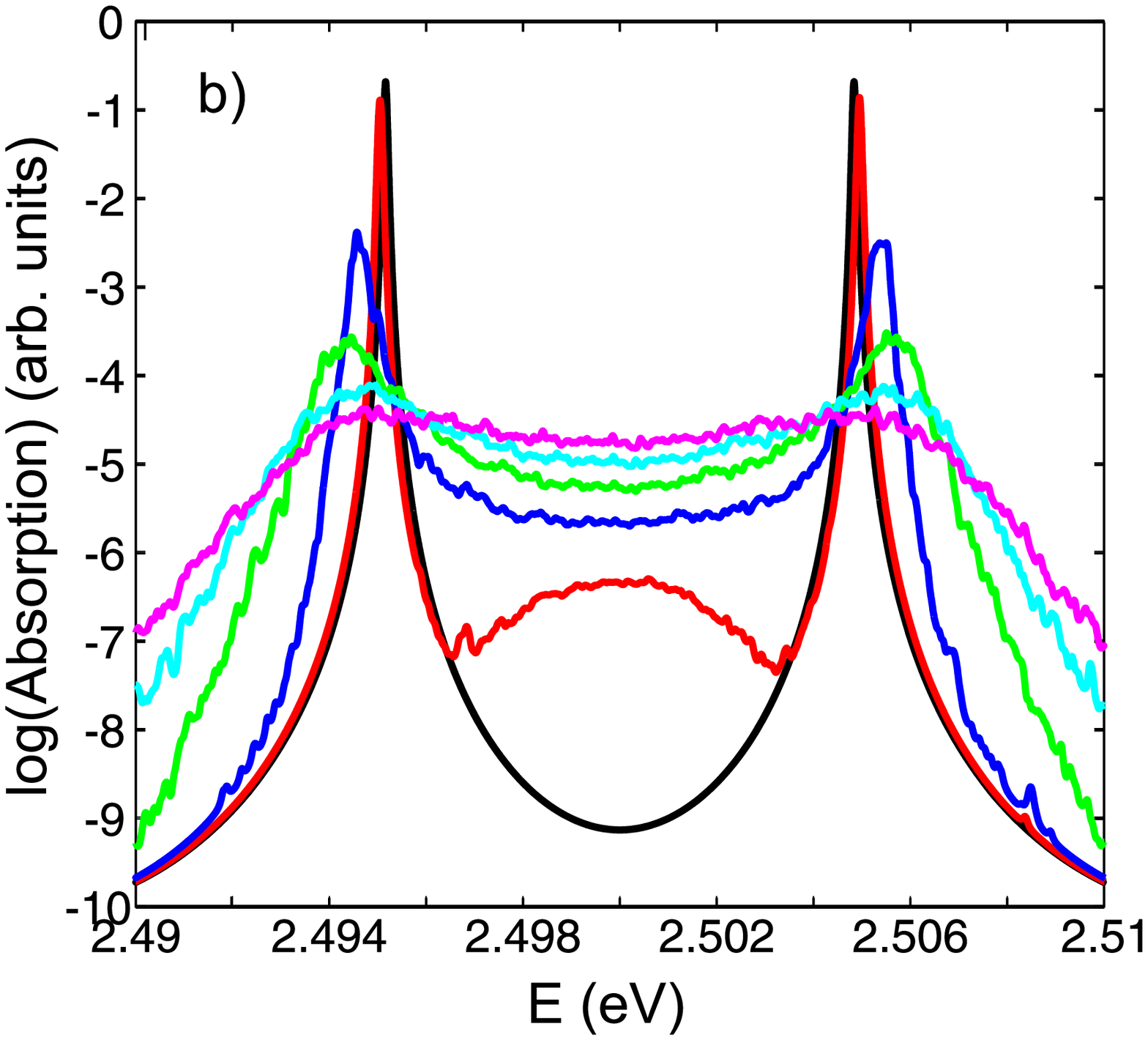}
\caption{(Color online) Polariton absorption spectrum in logarithmic scale for (i)
energy disorder with variance from 0 (black) to 5 meV (magenta) for polaritons at two high
symmetry points of the first Brillouin Zone: $M$ (a) and $\Gamma$ (b). A homogeneous Lorentzian
broadening of $\gamma$=50$\mu$eV was considered. Material parameters
correspond to ZnSe/CdSe systems, and can be found in
Ref. \onlinecite{KGP2007}.}
\label{figca}
\end{figure}
}

\newcommand{\figcb} {
\begin{figure}
\centering 
\includegraphics[width=0.3\textwidth]{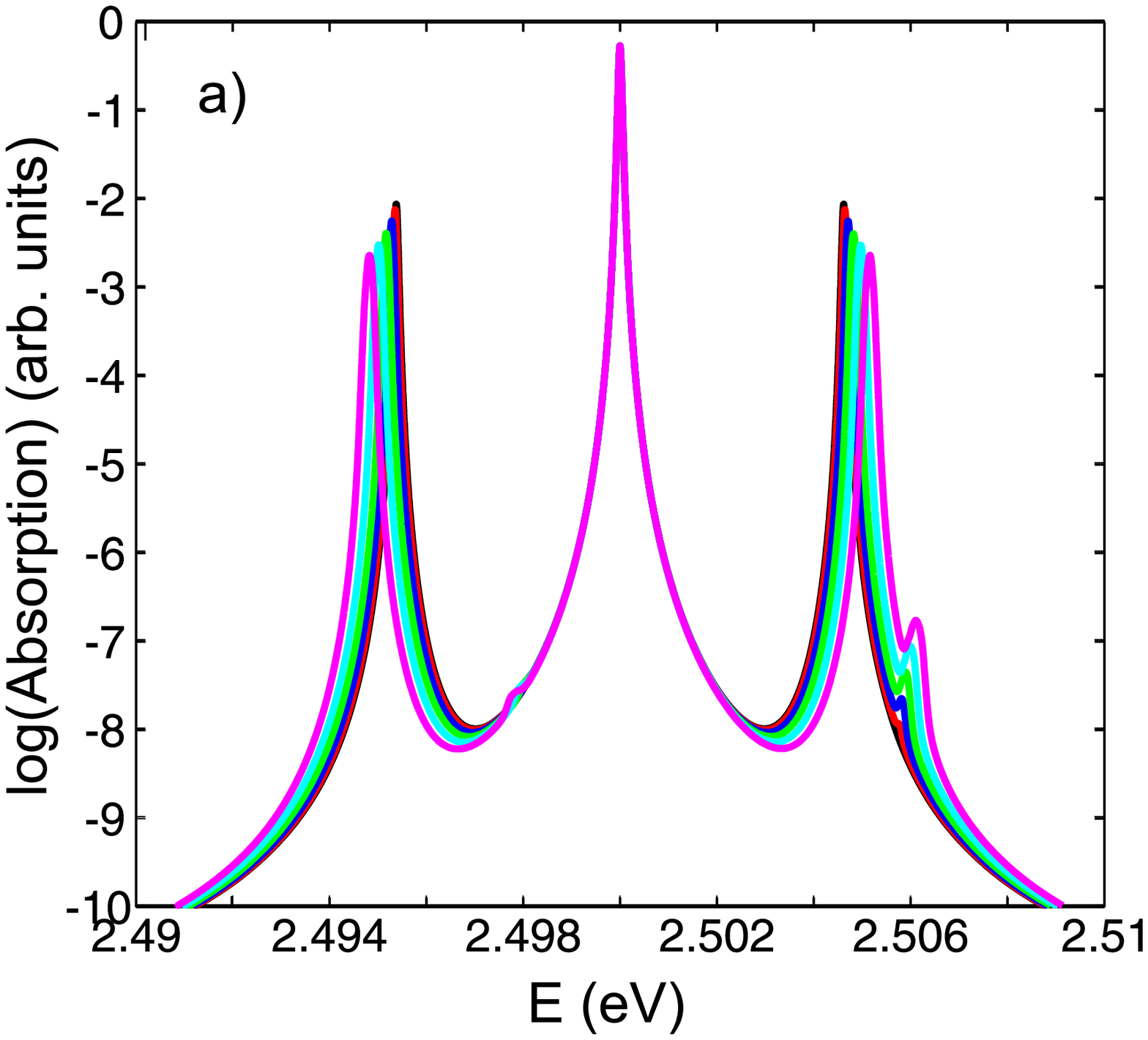}
\includegraphics[width=0.3\textwidth]{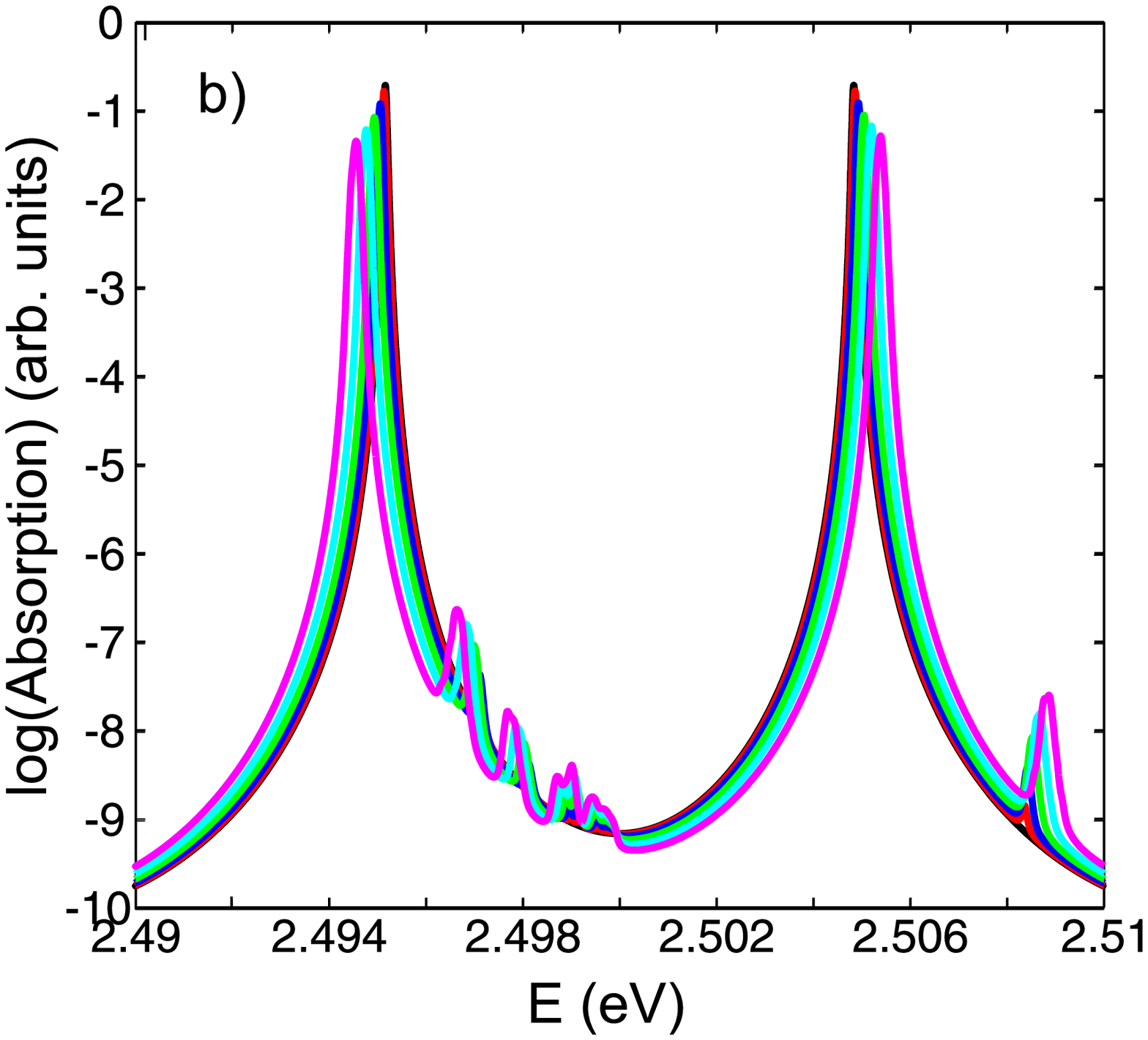}
\caption{(Color online) Polariton absorption spectrum in logarithmic scale for oscillator strength disorder with variance from 0 (black) to 50\% (magenta) of the average for polaritons at two high symmetry points of the first Brillouin Zone: $M$ (a) and $\Gamma$ (b). Other parameters as in Fig. \ref{figca}.}
\label{figcb}
\end{figure}
}

\newcommand{\figcc} {
\begin{figure}
\centering 
\includegraphics[width=0.3\textwidth]{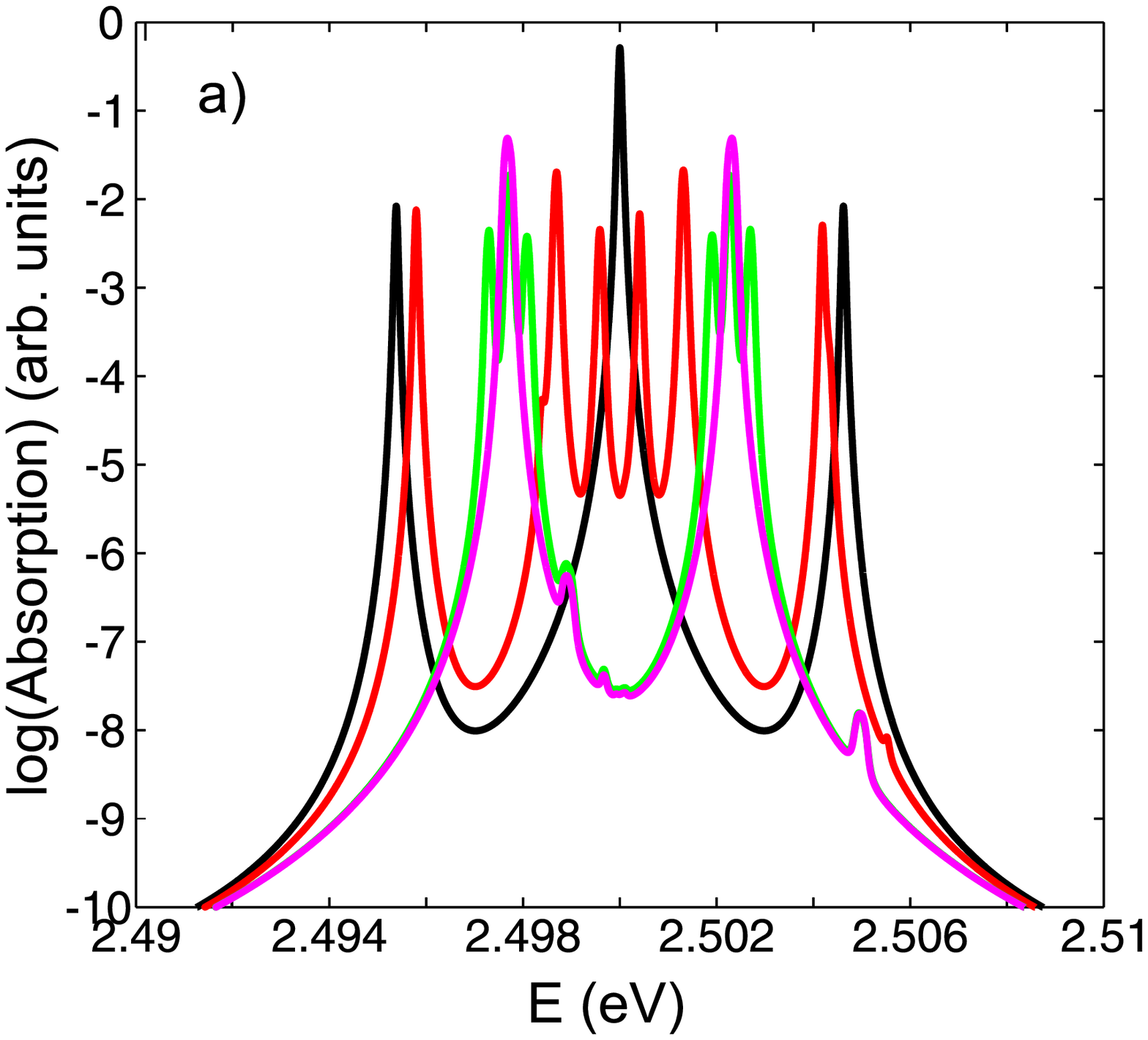}
\includegraphics[width=0.3\textwidth]{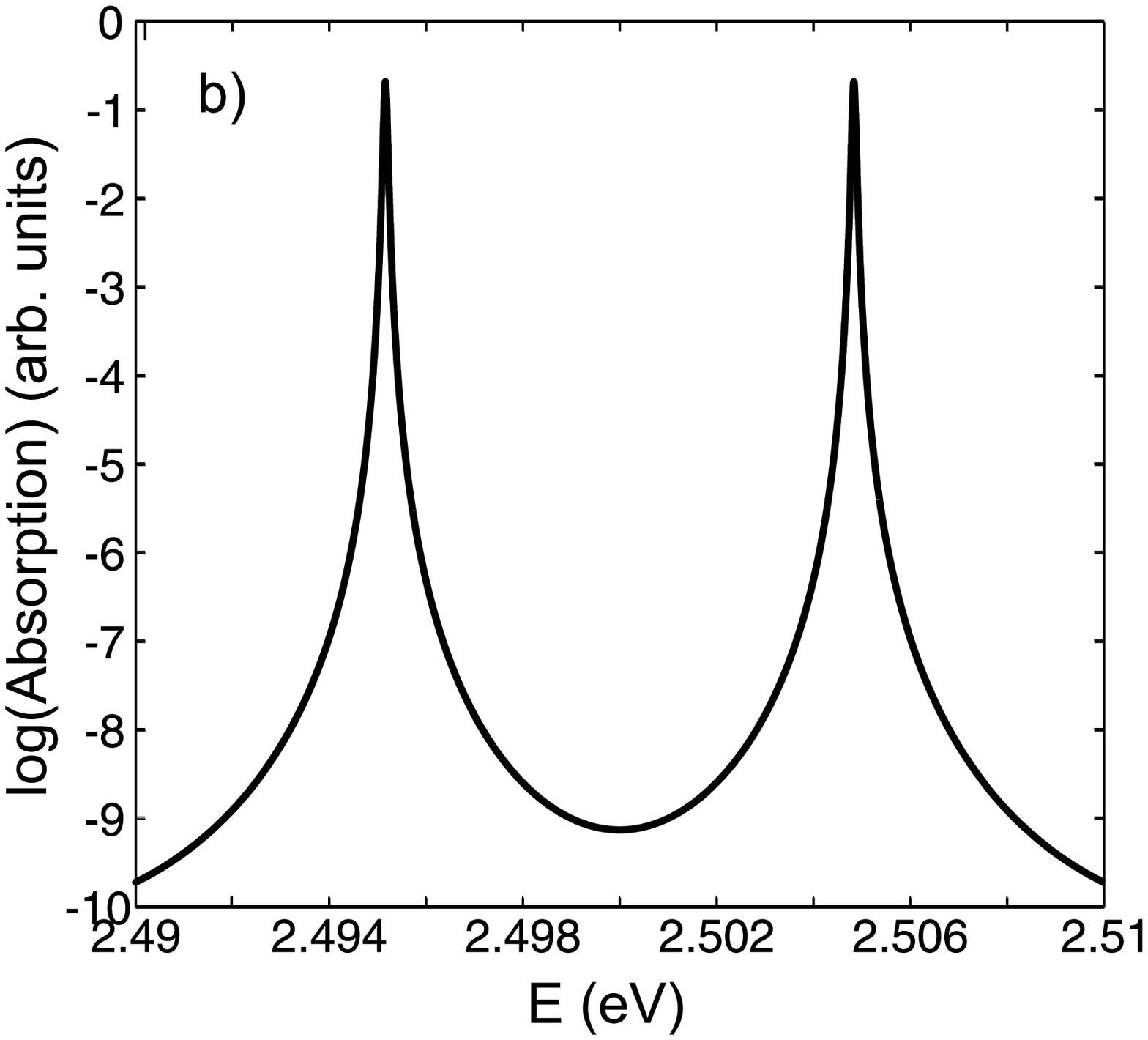}
\caption{(Color online) Polariton absorption spectrum in logarithmic scale for position disorder with variance from 0 (black) to 10\% (red), 30\% (green), and 50\% (magenta) of the lattice constant and for polaritons at two high symmetry points of the first Brillouin Zone: $M$ (a) and $\Gamma$ (b). Other parameters as in Fig. \ref{figca}.}
\label{figcc}
\end{figure}
}

\newcommand{\figd} {
\begin{figure}
\centering \includegraphics[width=0.3\textwidth]{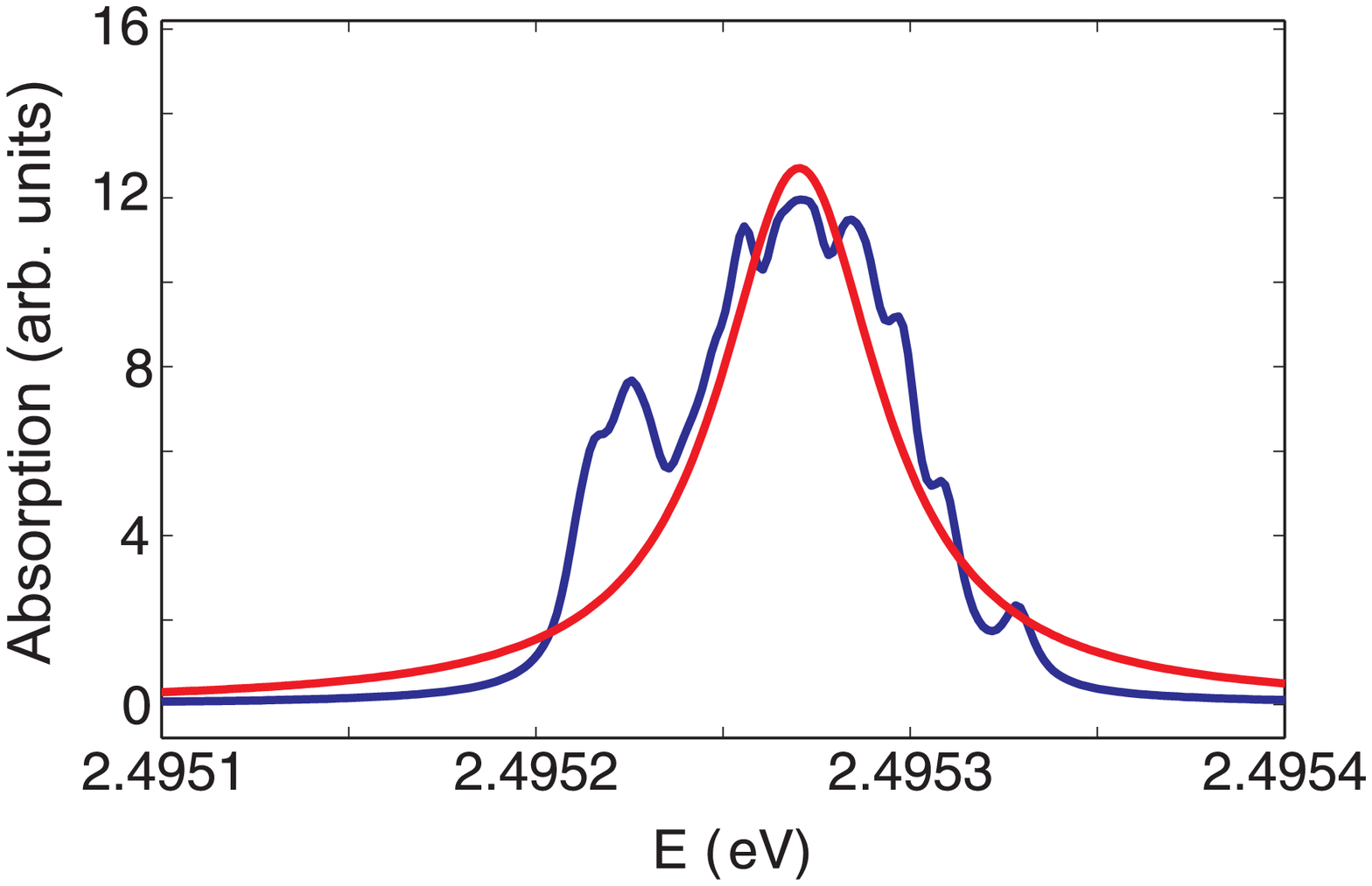}
\caption{(Color online) Absorption spectrum in the vicinity of the lower polariton
energy calculated exactly (blue) and within the perturbation theory
and broadened by Lorentzian (red) with $\gamma = 5 \mu eV$. 40
Realizations of energy disorder with variation $\sigma_\omega = 1$ meV
on the lattice with $20^2$ sites were considered.}
\label{figd}
\end{figure}
}

\begin{document}
\title{Microcavity polaritons in disordered exciton lattices}

\author{Michal Grochol and Carlo Piermarocchi}

\affiliation{Department of Physics and Astronomy, Michigan State
University, East Lansing, Michigan 48824 USA}

\date{\today}
\begin{abstract}
We investigate the interaction of excitons in a two dimensional
lattice and photons in a planar cavity in the presence of
disorder. The strong exciton-photon coupling is described in terms of
polariton quasi-particles, which are scattered by a
disorder potential. We consider three kinds of disorder: (a)
inhomogeneous exciton energy, (b) inhomogeneous exciton-photon
coupling and (c) deviations from an ideal lattice. These three types
of disorder are characteristic of different physical systems, and their
separate analysis gives insight on the competition between
randomness and light-matter coupling.  We consider conventional planar
polariton structures (with excitons resonant to photon modes
emitting normal to the cavity) and Bragg polariton structures, in
which excitons in a lattice are resonant with photon modes at a
finite angle satisfying the Bragg condition.  We calculate the
absorption spectra in the normal direction and at the Bragg angle by a
direct diagonalization of the exciton-photon Hamiltonian. We found
that in some cases weak disorder increases the light-matter coupling
and leads to a larger polariton splitting. Moreover, we found that the
coupling of excitons and photons is less sensitive to disorder of type
(b) and (c). This suggests that polaritonic structures realized with
impurities in a semiconductor or with atoms in optical lattices are
good candidate for the observation of some of the Bragg polariton
features.
\end{abstract}


\maketitle

\section{Introduction}
There is currently a considerable interest in the physics of strongly
coupled light-matter systems. Examples include experiments on the
Bose-Einstein condensation (BEC) of microcavity
polaritons,~\cite{deng03,kasprzak06,balili07,LKU+07} which are mixed
states of excitons and photons (see Refs. \onlinecite{savona99} and \onlinecite{keeling07}) and on the strong coupling of cavity photons and quantum
dot excitons.~\cite{peter05, reithmaier04, yoshie04} In order to combine the
properties of matter states in a lattice and planar photon modes, we
have recently investigated a structure consisting of an array of
quantum dots in a planar cavity,~\cite{KGP2007} as schematically shown
in Fig.~\ref{figa}. This particular geometry for light-matter
coupling can be realized in many different ways using either
semiconductor-based systems (quantum dots, impurities, metallic gates)
or atomic systems, such as atoms in optical lattices.  Experimental
and theoretical investigations show that impurity bound excitons can
have very small inhomogeneous broadening and strong confinement energy
(e.g. of the orders of $50$ meV~\cite{FKM2004, MBP+2006}). Moreover,
ordered arrays of single dopants,~\cite{SOK+2005} as well as the
control of a single impurity using scanning tunneling microscopy have
been experimentally demonstrated.\cite{RPR+2007, KKH+2007}

\figa Polaritons in structures similar to the one in Fig.~\ref{figa},
but involving photonic crystals\cite{gerace07} or optical
lattices\cite{CPS2007,BC2007} have been recently investigated
theoretically. Many of these structures have planar spatial periodicity,
which leads to very interesting and novel properties. We have recently
explored theoretically Bragg polariton modes at some special symmetry
points of the Brillouin zone boundaries.~\cite{KGP2007} These zone-edge
Bragg polaritons can have extremely small effective masses, typically
three orders of magnitude smaller than conventional cavity polaritons,
and behave effectively as Dirac quasiparticles, similar to light-mass (relativistic) electrons in graphene, which have been experimentally investigated recently.\cite{NGM+2005,NGM+2004}  Polaritons with Dirac dispersion could have interesting applications to the physics of polariton BEC
and superfluidity, as well as for spin-coupling
control.~\cite{QFP2006}

In Fig.~\ref{figa} we show disorder effects due to fluctuations in the
quantum dot size and position.  These disorder effects are the main
focus of this paper.  The effect of disorder on polaritons has been
extensively studied in the quantum well-microcavity case (see
Ref. \onlinecite{savona99} and references therein). It was found that
if the potential fluctuations due to the disorder are comparable with
the Rabi splitting then the two polaritonic peaks disappear and one
inhomogeneously broadened peak remains in the absorption
spectrum. Furthermore, a similar behavior has been theoretically
predicted for an ensemble of two-level systems coupled to a single
photon mode.~\cite{HSI1995} The dynamics of the wave packet and its localization in disordered one-dimensional cavity has been been studied too. \cite{AG2007} In this paper, we investigate the effect
on polariton modes of energy, oscillator strength, and position
fluctuations. The numerical results are discussed and compared to a
perturbation theory approach. The paper is organized as follows: The
theory is introduced in Sec. II, followed by results and discussion in
Sec. III. Conclusions are drawn in Sec. IV. 

\section{Theory}
We are going to investigate a disordered planar lattice embedded in a
planar microcavity structure. The role of the lattice is to localize
excitons and the physical mechanism of localization will not be
specified at this point. As discussed above, quantum dots, impurities,
metal gates and optical lattices can be used to localized the exciton
states. We further assume that only one excitonic level is present on
each site, implying a strong localization. This also allows us to
describe the exciton transfer process between sites using a
tight-binding approach. The case of quantum dots has been
extensively discussed in Ref. \onlinecite{KGP2007}. We start with the
Hamiltonian in second quantization ($\hbar = 1$)
\begin{eqnarray}
\nonumber \hat{H} &=& \sum_j \omega_j C^\dagger_j C_j + t_X \sum_{j, k
\in NB} C^\dagger_{j+k} C_j \\ &+& \sum_q \omega_q a^\dagger_q a_q +
\sum_{j q} \bigl ( i g_{j q} e^{i q R_j} a^\dagger_q C_j + h.c. \bigr
)~,
\end{eqnarray}
where $C^\dagger_j$ is the exciton creation operator on the $j^{th}$
site at position $R_j$ with energy $\omega_j$ and exciton-photon
coupling $g_{j q}$, $t_X$ is the inter-site energy transfer between
nearest neighbors sites, $a^\dagger_q$ is the creation operator
of the cavity photon mode with in-plane momentum $q$ and energy
$\omega_q$. By writing the coupling constant as $g_{j q} e^{i q R_j}$
we can analyze separately the disorder effect induced by position
fluctuations and oscillator strength fluctuations.

\subsection{Disorder properties}
We assume that the distance between localization sites is much larger
than the effective exciton localization length on each site. This justifies why fluctuations are site-uncorrelated.  For
energy fluctuations, defined as $\Delta \omega_j = \omega_j -
\omega_X$, $\omega_X$ being the average exciton energy, we have
\begin{eqnarray}
\langle \Delta \omega_j \rangle = 0
\end{eqnarray}
and
\begin{eqnarray} 
\quad \langle \Delta \omega_m
\Delta \omega_n \rangle = \sigma^2_\omega \delta_{mn}~,
\end{eqnarray}
where $\langle . \rangle$ indicates averaging over a Gaussian
ensemble. Similarly, we have for the oscillator strength disorder
\begin{eqnarray}
\langle g_{j q} \rangle = \langle g_{0 q} \rangle = \overline{g_{0
q}}, \quad \langle g_{k q} g_{l q'} \rangle - \langle g_{k q} \rangle
\langle g_{l q'} \rangle = \sigma^2_{qq'} \delta_{kl}~,
\end{eqnarray}
where we assume $g_{0q} = g e^{-q^2 \beta^2 / 4}$ with the characteristic size of the site\cite{KGP2007} $\beta$ and 
\begin{eqnarray}
\sigma^2_{qq'} = \sigma^2_g e^{-(q^2 + q'^2) \beta^2 / 4},  
\end{eqnarray}
where $\sigma^2_g = \langle g^2 \rangle - \langle g \rangle^2$ is the on-site oscillator strength variance at $q=0$. Finally, we obtain for the positional disorder 
\begin{eqnarray}
\nonumber \langle R_j \rangle &=& \overline{R_j}, \qquad \langle R_k R_l \rangle - \langle R_k \rangle \langle R_l \rangle = \sigma^2_R \delta_{kl},\\
\langle e^{i q R_j} \rangle &=& e^{i q \overline{R_j}} e^{-q^2 \sigma^2_R / 2},
\end{eqnarray}
where the positions $\overline{R_j}$ identify the ideal two
dimensional lattice. Furthermore, we assume for simplicity that the different kinds of disorder are uncorrelated. This allows us to treat them separately in the perturbation approach described below.

\subsection{Polariton scattering}
We can separate the fluctuation-independent part of the Hamiltonian by
defining the Fourier transform of the exciton operators as
\begin{eqnarray}
\nonumber C_j^\dagger &=& \frac{1}{\sqrt{N}} \sum_{q \in 1.BZ} e^{i q
\overline{R_j}} C_q^\dagger, \\ C_j &=& \frac{1}{\sqrt{N}} \sum_{q \in
1.BZ} e^{-i q \overline{R_j}} C_q,
\end{eqnarray}
where $\overline{R_j}$ corresponds to the $j^{th}$ site of a two
dimensional ideal lattice and $N$ is the total number of sites.  The sum
over $q$ states is restricted to the first Brillouin Zone (BZ) of the
reciprocal lattice space.  This fluctuation-independent Hamiltonian
then takes the form
\begin{eqnarray}
\label{regular}
\nonumber \hat{H}_0 &=& \sum_q \Bigl \{ \omega_{Xq} C^\dagger_q C_q +
\sum_{Q} \omega_{q+Q} a^\dagger_{q+Q} a_{q+Q} \\ &+& \sum_{Q}
(\tilde{g}_{q+Q} a^\dagger_{q+Q} C_q + h.c.) \Bigr \},
\end{eqnarray}
where $Q$ is a reciprocal lattice vector, $\tilde{g}_{q} = \sqrt{N}
g_{0 q}$ is the renormalized coupling constant, and the exciton
dispersion of a square lattice reads
\begin{eqnarray}
\omega_{Xq} &=& \omega_X - 2 t_X \left ( \cos(q_x a) + \cos(q_y a) \right ),
\end{eqnarray}
where $a$ is the lattice constant. We can then define the disorder coupling constants
\begin{eqnarray}
\nonumber \Delta\omega_q &=& \frac{1}{\sqrt{N}} \sum_j e^{i q
\overline{R_j}} \Delta \omega_j,\\ \nonumber \eta_{qq'}^o &=&
\frac{1}{\sqrt{N}} \sum_j e^{i (q-q') \overline{R_j}} (g_{j q} -
\overline{g}_{0 q}), \\ \eta_{qq'}^p &=& \frac{1}{\sqrt{N}} \sum_j
(e^{i q R_j} - e^{i q \overline{R_j}}) e^{-i q' \overline{R_j}}
\overline{g}_{0 q},
\end{eqnarray}
where the index $o$ ($p$) indicates oscillator strength (position) disorder.
The statistical properties of these functions are determined by their average value as
\begin{eqnarray}
\label{disprop}
\nonumber \langle \Delta \omega_k \rangle &=& 0, \\
\nonumber \langle \eta^o_{qq'} \rangle &=& 0,  \\ 
\langle \eta^p_{qq'} \rangle &=& \sum_K \delta_{q,q'+K} \xi_q, 
\end{eqnarray}
and correlations according to
\begin{eqnarray}
\label{correlations}
\nonumber \langle \Delta \omega_k \Delta \omega_l^* \rangle &=& \sigma^2_\omega \delta_{kl}, \\
\langle \eta^o_{qq'} \eta^{*o}_{kk'} \rangle &=& \sum_K \delta_{q-k,q'-k'+K} \sigma_{qk}, \\
\nonumber \langle \eta^p_{qq'} \eta^{*p}_{kk'} \rangle &-& \langle \eta^p_{qq'} \rangle \langle \eta^{*p}_{kk'} \rangle = \sum_K
\delta_{q-k,q'-k'+K} \zeta_{qk},
\end{eqnarray}
where $K$ is a reciprocal lattice vector and 
\begin{eqnarray}
\xi_{q} &=& \tilde{g}_{0 q} (e^{-q^2 \sigma^2_R / 2} - 1),\\
\nonumber \zeta_{qk} &=& \overline{g}_{0 q}
\overline{g}_{0 k}^* (e^{-|q+k|^2 \sigma^2_R / 2} - e^{-q^2 \sigma^2_R
/ 2} e^{-k^2 \sigma^2_R / 2}).
\end{eqnarray}
The disorder terms for the three different mechanisms can then be written in a compact form as
\begin{eqnarray}
\label{disham}
\nonumber \hat{H}^e &=& \frac{1}{\sqrt{N}} \sum_{q k} \Delta \omega_k
C^\dagger_{q+k} C_q, \\
 \hat{H}^{o(p)} &= & \sum_{qq'} (\eta_{qq'}^{o(p)}
a^\dagger_q C_{q'} + h.c.).
\end{eqnarray}
We can find the eigenvalues and eigenvectors of the disorder free Hamiltonian by solving
\begin{eqnarray}
\label{wavefun}
\hat{H}_0 |P_{nq} \rangle = \Lambda_{nq} |P_{nq} \rangle~.
\end{eqnarray}
These are the disorder-free polariton states that can be written in the form
\begin{eqnarray}
|P_{nq} \rangle &=& P_{nq}^\dagger |0 \rangle = (u_{nq}
C_q^\dagger + \sum_Q v_{n q+Q} a_{q+Q}^\dagger) |0 \rangle,
\end{eqnarray}
where $n$ is a band index, $v_{nq}$ and $u_{nq}$ are Hopfield
coefficients,\cite{Hopfield1958} and $| 0 \rangle$ is the exciton-photon vacuum. We can use these states and obtain an
effective disorder potential for polaritons as
\begin{eqnarray}
\label{scatpot}
\nonumber V^{J}_{nn'qq'} &=& \langle P_{nq} |\hat{H}_{1(2)}|
P_{n'q'} \rangle, 
\end{eqnarray}
where $J \in \{ e, o, p \}$ labels energy, oscillator strength, and positional disorder, respectively.
This gives for the energy disorder
\begin{eqnarray}
V^e_{nn'qq'} &=&
\frac{\omega_{q-q'}}{\sqrt{N}} u^*_{nq} u_{n'q'},
\end{eqnarray}
and for the oscillator strength  (positional) disorder
\begin{eqnarray}
\nonumber V^{o(p)}_{nn'qq'} &=& \sum_Q \left ( \eta^{o(p)}_{q+Q q'} u_{n'q'} v^*_{n
q+Q} + \eta^{* o(p)}_{q+Q q'} u^*_{nq} v^*_{n' q'+Q} \right )~.  \\
\end{eqnarray}

\subsection{Absorption spectrum}
The absorption spectrum at an angle determined by the inplane $q$ of the
cavity photon can be calculated using the full propagator of the
disordered system, which can be written in terms of the disorder-free
polariton states as
\begin{eqnarray}
\hat{G}(\omega) &=& \sum_{n n' kk'} | P_{nk} \rangle G_{nn'kk'}(\omega)
\langle P_{n'k'} | 
\end{eqnarray}
with
\begin{eqnarray}
G_{nn'kk'}(\omega) &=&
G^0_{nk}(\omega) \delta_{nn'} \delta_{kk'} \nonumber \\ &+& G^0_{nk}(\omega)
T_{nn'kk'}(\omega) G^0_{n'k'}(\omega),
\end{eqnarray}
where 
\begin{eqnarray}
G^0_{nk}(\omega) = \frac{1}{\omega - \Lambda_{nk}+ i \epsilon} 
\end{eqnarray}
 is the
Green's function for disorder-free polaritons. $T$ is the scattering
T-matrix that can be expressed as
\begin{eqnarray}
\nonumber T(\omega) &=& \sum_j T^{(j)}(\omega), \\
T^{(j)}_{nn'kk'}(\omega) &=& \langle P_{nk} | \hat{V}
(\hat{G}^0(\omega) \hat{V})^j | P_{n'k'} \rangle.
\end{eqnarray}
The imaginary part of the polariton propagator projected on a photon
mode at a given wavevector $| \gamma \rangle \equiv | k_\gamma \rangle$ gives the absorption spectrum at the
corresponding emission angle as
\begin{eqnarray}
A_\gamma(\omega) &=& -\mbox{Im} \, G_\gamma(\omega)~, 
\end{eqnarray}
with
\begin{eqnarray}
\nonumber G_\gamma(\omega) &=& \langle \gamma | \hat{G}(\omega) | \gamma \rangle
= \sum_{n n'} v_{n k_\gamma} v^*_{n' k_\gamma} G_{nn'k_\gamma
k_\gamma}(\omega).
\end{eqnarray}
We can explicitely average over the disorder configurations to obtain 
\begin{eqnarray}
\nonumber \langle G_\gamma(\omega) \rangle &=& \sum_{n n'} v_{n
k_\gamma} v^*_{n' k_\gamma} \langle G_{nn'k_\gamma k_\gamma}(\omega)
\rangle 
\end{eqnarray}
with
\begin{eqnarray}
\langle G_{nn'k_\gamma k_\gamma}(\omega) \rangle
&=& G^0_{nk}(\omega) \delta_{nn'} \delta_{kk'} \\ &+&
\nonumber  G^0_{nk_\gamma}(\omega) \langle T_{nn'k_\gamma k_\gamma}(\omega)
\rangle G^0_{n'k_\gamma}(\omega).
\end{eqnarray}
In the case of disorder due to energy inhomogeneity, the ensemble
averaging leads to the conservation of the momentum, $\langle
T_{nn'k k'}(\omega) \rangle = \langle T_{nn'k k}(\omega) \rangle
\delta_{kk'}$, since the averaging selects only diagonal terms in Eq.~(\ref{correlations}). We note that this is the same as in the case of one particle scattering on an impurity ensemble.\cite{Mahan} The diagonal elements of the $T$-matrix can be
interpreted as an effective self-energy, $\Sigma_{nk}(\omega) =
T_{nnkk}(\omega)$ and within the pole approximation
$\Sigma_{nk}(\omega = \Lambda_{nk})$, their real part gives a shift of
the polariton levels.  The shift can also be calculated using a direct
perturbation theory approach as shown in the following section. 

\begin{widetext}
\subsection{Perturbation theory}
Here we calculate the corrections to the energy up to the second order in
the disordered potential, using the statistical properties of the
disorder derived above. We start with the first order energy shift for
all three kinds of disorder
\begin{eqnarray}
E^{1,e}_{nk} &=& \frac{\Delta \omega_0}{\sqrt{N}}
|u_{nk}|^2, \qquad E^{1,o(p)}_{nk} = 2 \mbox{Re} \left \{
\sum_K \eta_{k+K k}^{o(p)} u_{nk} v^*_{n k+K} \right \},
\end{eqnarray}
where $u_{nk}$ and $v^*_{n k+K}$ are defined in
Eq. (\ref{disham}). The average value of the energy shift
\begin{eqnarray}
\label{enave}
\langle E^{1,e}_{nk} \rangle &=& 0, \\
\label{osave}
\langle E^{1,o}_{nk} \rangle &=& 0, \\
\label{posavea}
\langle E^{1,p}_{nk} \rangle &=& 2 \mbox{Re} \left \{ \sum_K
\xi_{k+K} u_{nk} v^*_{n k+K} \right \},
\end{eqnarray}
and variances are
\begin{eqnarray}
\label{envar}
\langle (E^{1,e}_{nk})^2 \rangle &=& \frac{\sigma^2_\omega}{N} |u_{nk}|^4; \\
\label{osvar}
\langle (E^{1,o}_{nk})^2 \rangle &=& 4 \mbox{Re} \left \{ \sum_{K, K'} \sigma_{k+K,k+K'}|u_{nk}|^2 v^*_{n k+K} v^*_{n k+K'} \right \},\\
\label{posvar}
\langle (E^{1,p}_{nk})^2 \rangle - \langle E^{1,p}_{nk} \rangle^2 &=& 4 \mbox{Re} \left \{ \sum_{K, K'} \zeta_{k+K k+K'} |u_{nk}|^2
v^*_{n k+K} v^*_{n k+K'} \right \}.
\end{eqnarray}
The second order energy shift in the {\it non-degenerate} case is given by
\begin{eqnarray}
E^{2,e}_{nk} &=& \frac{1}{N} \sum_{n'k'} \frac{|\Delta
\omega_{k-k'}|^2 |u_{nk}|^2 |u_{n'k'}|^2}{\Lambda_{nk} -
\Lambda_{n'k'}}, \\ 
E^{2,o(p)}_{nk} &=& \sum_{n'k'} \frac{|
\sum_K (\eta^{o(p)}_{k+K k'} u_{n'k'} v^*_{n k+K} + \eta^{* o(p)}_{k+K
k'} u^*_{nk} v^*_{n' k'+K})|^2}{\Lambda_{nk} - \Lambda_{n'k'}}.
\end{eqnarray}
In the case of energy disorder, average and variance can be
calculated as
\begin{eqnarray}
\label{enave2}
\langle E^{2,e}_{nk} \rangle &=& \frac{\sigma^2_\omega}{N} \sum_{n'k'}
\frac{|u_{nk}|^2 |u_{n'k'}|^2}{\Lambda_{nk} - \Lambda_{n'k'}}, \\
\label{envar2}
\langle (E^{2,e}_{nk})^2 \rangle - \langle E^{2,e}_{nk} \rangle^2 &=&
\frac{\sigma^4_\omega}{N^2} \sum_{n'k'} \frac{|u_{nk}|^4
|u_{n'k'}|^4}{(\Lambda_{nk} - \Lambda_{n'k'})^2}.
\end{eqnarray}
The average value for the the oscillator strength and positional
disorder are given by
\begin{eqnarray}
\label{osave2}
\langle E^{2,o(p)}_{nk} \rangle &=& \sum_{n'k'} \frac{\chi^{o
(p)}_{nkn'k'}}{\Lambda_{nk} - \Lambda_{n'k'}},
\end{eqnarray}
where
\begin{eqnarray}
\nonumber \chi^{o (p)}_{nkn'k'} &=& \sum_{K K'} \left (
\vartheta^{1 o(p)}_{kk'KK'} |u_{n'k'}|^2 v^*_{n k+K} v^*_{n k+K'} +
\vartheta^{2 o(p)}_{kk'KK'} u_{nk} u_{n'k'} v^*_{n k+K} v^*_{n' k'+K'}
\right. \\ 
&+& \left. \vartheta^{* 2 o(p)}_{k'kKK'} u^*_{nk}
u^*_{n'k'} v_{n k+K'} v_{n' k'+K} + \vartheta^{1 o(p)}_{k'kKK'}
|u_{nk}|^2 v_{n' k'+K} v_{n' k'+K'} \right ),\\
\nonumber \vartheta^{1 o(p)}_{kk'KK'} &=& \langle \eta^{o (p)}_{k+K
k'} \eta^{* o(p)}_{k+K' k'} \rangle, \\
\nonumber \vartheta^{2 o(p)}_{kk'KK'} &=& \langle \eta^{o (p)}_{k+K k'} \eta^{o(p)}_{k'+K' k}
\rangle.
\end{eqnarray}
Expressions for the variance are not shown but their calculation is straigthforward. The degenerate case for the second order shift will be discussed in the next section.

\end{widetext}

\section{Results and discussion}

Here we present the absorption spectra calculated numerically
for various strengths of disorder. We consider the case of an
excitation at the $M$ point (finite excitation angle corresponding to
$q_M=(\pi/a,\pi/a)$) and at the $\Gamma$ point (normal excitation). The exciton energy is tuned into resonance with the corresponding cavity modes. We have used material parameters of CdSe/ZnSe systems as summarized in Table III of
Ref. \onlinecite{KGP2007}. The full Hamiltonian corresponding to a
lattice with $31^2$ sites and photon modes in an energy window
$[0.9\omega_X, 1.1\omega_X]$ is diagonalized. In order to make the
results independent of the system size we keep fixed the total
disorder-free coupling constant $\tilde{g}_{q}$ and we use periodic
boundary conditions. 
\figca

Let us start with the effect of the energy
disorder in the absorption spectra as shown in Fig.~\ref{figca}. The disorder-free case (black line) shows lower (LP) and upper (UP)
polariton peaks with half-exciton and half-photon
character. Additionally, in the $M$-point case (a), there is a central
purely photonic peak. The polariton splitting, i.e. the energy
difference between the lower and the upper polariton energy, is larger
by $\sim 0.5$ meV at the $\Gamma$-point than at the $M$-point due to
the stronger coupling in the normal direction.~\cite{KGP2007} The disorder shifts the LP (UP) energy: According to the perturbation theory the first order shift of the LP (UP) is zero (as seen from Eq.~(\ref{enave})) and the second order
shift is negative (positive), according to Eq. (\ref{enave2}), and
leads to an {\it increase} of the polariton splitting. Therefore, a weak
disorder can increase the robustness and visibility of the polariton
splitting with respect to the ideal case. The fact that a weak exciton
disorder increases the observability of a coherent effect may be seen as 
counter-intuitive, but it has been also found in similar contexts,
e.g. in the exciton Aharonov-Bohm effect.\cite{GGZ2006,GZ2007} Additionally, by increasing disorder the UP and LP peaks broaden. Within the perturbative approach we find that the broadening of the polariton modes is equal to $\frac{\sigma_\omega}{\sqrt{N}}$ in the lowest order (see  Eq.~(\ref{envar})), the second order being proportional to $\frac{\sigma^2_\omega}{N}$ (see Eq.~(\ref{envar2})). Moreover, in order to check our perturbative approach we have also performed a direct comparison between the full numerical solution and the analytical result from perturbation theory for weak disorder, shown in Fig. \ref{figd} and an excellent agreement is found.

\figd

Notice that for both $\Gamma$ and $M$ polaritons,
a new central very broad excitonic peak appears. However, in the $M$-point
case a central peak corresponding to a pure photonic mode, clearly
insensitive to the disorder, is superposed to the broad central excitonic
background.  As the width of the energy distribution due to the
inhomogeneous energy disorder approaches the value of the lower and upper
polariton splitting, the three peaks merge and the spectrum consists
of a central narrow (photon-like) peak and a broad excitonic
background.

\figcb

From an experimental point of view, state-of-the-art quantum dot
samples have inhomogeneous broadening of several meV, which is comparable to typical values of the polariton splitting.
This makes the observation of the polariton splitting challenging. On the other hand,
impurities have very narrow energy distribution and the planar lattice
can be realized by patterned ion-implantation methods.  Array of
impurties with deep excitonic levels could be obtained by this
method. Assuming deep energy levels, Frenkel-like excitons bound to impurities will have very small inhomogeneous energy
broadening. Nevertheless, using ion implantation, it is hard to control the number of impurities on every lattice site. Thus, the lattice of impurities has a very narrow on-site energy distribution but large oscillator strength disorder due to impurity number fluctuation. The same kind of disorder may occur in atoms trapped in optical lattices. In order to investigate this effect, we consider a Gaussian distribution of the exciton oscillator strength in the coupling constants $g_{jq}$. A Poisson distribution is another possibility but we have not investigated that case. The absorption spectra for different oscillator strength disorder are shown in Fig.~\ref{figcb}. The behaviour of the LP and UP peaks for both $\Gamma$ and $M$-point is similar to the case of the energy disorder. In particular, we observe a decrease (increase) of LP (UP) energy, which is again consistent with the perturbation theory as seen in Eqs. (\ref{osave}) and (\ref{osave2}). We also remark that even for very large fluctuations of the oscillator strength both the broadening and energy shift remain smaller with repect to the inhomogeneous energy disorder.

\figcc

Finally, we have also investigated absorption spectra in the presence
of positional disorder, i.e. deviations in the exciton localization
sites from the ideal lattice, as shown in Fig.~\ref{figcc}. Clearly,
this disorder plays a role only for $M$-point polaritons, since the
$\Gamma$-point does not probe the lattice symmetry (Fig.~\ref{figcc}b).  In contrast to
the oscillator strength disorder and to the energy disorder,
positional disorder leads to a blueshift (redshift) of the lower
(upper) polariton peak, which implies a reduction of the polariton
splitting. This is related to the fact that the first order
perturbation theory is non-zero in this case, as seen in
Eq. (\ref{posavea}). The behaviour of the central polariton (CP) peak is more complex. First, it splits into
three lines for weak disorder. Then the remaining central peak splits again into a doublet for stronger disorder. The first splitting can be understood easily taking into account CP's three-fold degeneracy\cite{KGP2007} at $q_M$ which is lifted in presence of disorder and consequently one energy shifts upwards (downwards) and one stays in the middle. With increasing strength of disorder, the state with $q_M$ mixes with neighboring $q$ and doublet appears.

Moreover, assuming zero exciton transfer, no anylitical expressions can be derived because the CP is $N$-fold degenerate (in the BZ). On the other hand, assuming a finite exciton transfer rate $t_X \neq 0$, the $N$-fold degeneracy decreases to a three-fold degeneracy at the $M$-point. Since these levels are purely photonic, the first order matrix elements are zero and second order perturbation theory has to be used. The matrix elements of the second order effective Hamiltonian read
\begin{eqnarray}
\Theta_{mn}^{o(p)} &=&   \kappa_{mn}^{o(p)} \sum_{n'k'} \frac{|u_{n'k'}|^2}{\Lambda_{CP k_M} - \Lambda_{n'k'}}, 
\end{eqnarray}
where
\begin{eqnarray}
\kappa^{o (p)}_{mn} &=& \sum_{KK'}  \eta^{o (p)}_{k+K k'} \eta^{* o(p)}_{k+K' k'} v^*_{m k+K} v_{n k+K'}.
\end{eqnarray}
The correction to the unperturbed energy is found by averaging the eigenvalues of $\Theta$, whichs cannot be done analytically. A good approximation consists in averaging first the matrix elements 
\begin{eqnarray}
\label{cendeg}
\langle \kappa^{o (p)}_{mn} \rangle = \vartheta^{1 o(p)}_{kk'KK'} v^*_{m k+K} v_{n k+K'}
\end{eqnarray}
before calculating the eigenvalues. We have shown that this approximation introduces errors of the fourth order in the disorder potential and gives good agreement with the exact numerical calculation.

If the positional disorder is of the same order of the lattice constant, i.e. for $\sigma_R \sim 0.5a$, the lattice becomes close to a set of random sites. In this limit, instead of averaging the eigenvalues, the random phase $e^{i q R_j}$ in the coupling constants can be averaged first, which leads to a position independent value $\langle g e^{i q R_j} \rangle = g$. This results in a $\Gamma$-point like spectrum with an upper and lower polariton as clearly shown for the strongest disorder in Fig.~\ref{figcc}a (magenta).

\section{Conclusions}

We have investigated numerically and analytically the role of disorder in structures consisting of a lattice of dots or impurities embedded in a planar microcavity.  We have focused on Bragg polaritons occurring at the Brillouin zone boundary and we have shown that small energy disorder, i.e. diagonal disorder, with respect to the light-matter coupling is necessary to observe polariton features in the absorption spectra. Moreover, we have found that polariton is very robust against oscillator strength and position disorder. Our results show that a lattice consisting of deep impurities with Frenkel excitons or atoms in an optical lattice structure are promising candidates for the experimental observation of the Bragg polaritons.

\begin{acknowledgments}
This research was supported by the National Science Foundation, Grant No. DMR-0608501.
\end{acknowledgments}

\bibliographystyle{apsrev}

\end{document}